\documentclass[12pt,a4paper]{article}
\usepackage[utf8]{inputenc}
\usepackage{xcolor}
\usepackage[english]{babel}
\usepackage{amsmath}

\usepackage{amsfonts}
\usepackage{upgreek}
\usepackage{amssymb}
\usepackage{graphicx}
\usepackage{caption}
\usepackage{mathtools}
\usepackage{centernot}

\usepackage[font=footnotesize,labelfont=bf]{caption}
\usepackage[font=footnotesize]{subcaption}
\usepackage[left=3.5cm,right=3.5cm,top=3.8cm,bottom=3.8cm]{geometry}
\usepackage{dcolumn}
\usepackage{bm}
\usepackage{booktabs}
\usepackage{multirow}
\usepackage{framed}
\usepackage{enumerate}
\usepackage{algpseudocode}
\usepackage{algorithm}
\usepackage[nottoc]{tocbibind}
\usepackage{siunitx}

\newcolumntype{M}[1]{>{\centering\arraybackslash}m{#1}}
\newcolumntype{P}[1]{>{\centering\arraybackslash}p{#1}}

\usepackage[square,numbers]{natbib}
\usepackage[breaklinks]{hyperref}
\usepackage{cleveref}
\usepackage{ntheorem}
\theoremstyle{break}

\newtheorem{theorem}{Theorem}
\newtheorem{assumption}{Assumption}

\usepackage{gensymb}
\usepackage[runin]{abstract}

\title{Mathematical certification of motion planning on uncertain terrain with limited perception: a case study}
\author{Nikolaos Skouloudis and Alexandre Megretski }
\date{}
\begin{document}
\renewcommand{\abstractname}{\normalfont\scshape{Abstract}.}
\maketitle
\begin{abstract}
   We design a controller for an agent whose mission is to reach a stationary target while avoiding a family of obstacles which are not known a-priori. The agent moves in the two dimensional plane with non-trivial double integrator dynamics and receives only local information from its surroundings. Under mild assumptions on the family of obstacles (smoothness, sufficient distance from each other, bounded curvature, etc), we prove that our control algorithm yields guaranteed obstacle avoidance and convergence to the target.  
\end{abstract}
\noindent 
\section{Problem Setup}
This article presents a case study of an agent moving in a two dimensional plane filled with multiple stationary obstacles (not known a-priori) with the aim of reaching a known fixed target while avoiding collisions with the obstacles. The target is located at the origin and we consider a family of non-intersecting sets (obstacles) in the two dimensional plane. The boundary of each obstacle is assumed to be a smooth curve of bounded curvature, without self-intersections. Equivalently, the boundary of the i-th obstacle is the set $\left\{f_i(s): s\in[0,1]\right\}$, where $f_i:[0,1]\to\mathbb{R}^2$ and satisfies
\begin{eqnarray}
    \left \lvert \dot{f}_i\right\rvert&=&v_0, \nonumber  \\
     \ddot{f}_i&=&\frac{v_0w_i}{r_0}S\dot{f}_i.
\end{eqnarray}
Here $v_0$ and $r_0$ are the fixed growth-rate and the minimum radius of curvature of any obstacle boundary curve, $\lvert w_i \rvert \leq 1$ is relative curvature of the $i$th boundary curve and $S$ is a rotation matrix by $\frac{\pi}{2}$. We also assume that the obstacles are sufficiently far away from each other and that the boundary curve of each obstacle allows sharp turning (see \S\ref{sec:locking}), namely,
\begin{assumption}
\begin{enumerate}[(i)]
    \item $d_H\left(f_i\left([0,1]\right),f_j\left([0,1]\right)\right)\geq\frac{8}{5}r_0$ for all $i,j$,
    \item For every $p\in f_i\left([0,1]\right)$ we have $B(p+r_0n_p,r_0)\bigcap f_i\left([0,1]\right)=\emptyset$,
\end{enumerate}
\end{assumption}
where $d_H\left(X,Y\right)$ denotes the Hausdorff distance between $X \subset \mathbb{R}^2$ and $Y \subset \mathbb{R}^2$, $B(x,r)$ is the open ball centred at $x\in \mathbb{R}^2$ with radius $r>0$ and $n_p$ is the outward normal unit vector to the obstacle boundary curve at $p\in\mathbb{R}^2$. We stress that condition (ii) can be dropped when the obstacles are convex.\\
On the other had, we assume that the agent has double integrator dynamics,
\begin{equation}
\begin{bmatrix}
     \dot{x}\\\dot{v}
\end{bmatrix}\in\begin{bmatrix}
          v \\
          \frac{a_0u}{v_0}Sv
     \end{bmatrix},
     \label{eq:agent_dynamics}
\end{equation}
where $(x,v)\in \mathbb{R}^4$ denotes the agent's position and velocity, $a_0=\frac{v_0^2}{r_0}$ is the nominal acceleration of the agent and $u \in \mathbb{R}$ is the control action. We also define $r_a$ to be the minimum turning radius of the agent and consider bounded actuation $\lvert u \rvert \leq M$ where $M=\frac{r_0}{r_a}:=5$ (see \S \ref{sec:tracking} for a justification in selecting the latter value). Note that the agent moves at a constant velocity $\lvert v \rvert =v_0$ which concurs with the growth-rate of the obstacle boundary curve, however this does not pose significant restrictions in the
family of obstacles that are considered as it is not an intrinsic curve property (reparametrization). With no loss in generality, we allow the initial velocity, $v(0)$, to point in the direction of the origin, i.e. $v(0)=-v_0\frac{x(0)}{\left \lvert x(0) \right \rvert}$. We allow discontinuous and causal control with multiple but finite mode switches, thus existence and well posedness of solutions to (\ref{eq:agent_dynamics}) should be verified (see \S\ref{sec:control}).  \\
Moreover, we assume that the agent receives information about its surroundings via a depth sensor, which is modelled as a vector of length $d_m=\frac{4}{5}r_0$ starting from the agent's location and oriented at an angle $\phi$ from the agent's velocity vector. In proximity of an obstacle, the sensor provides: (i) $d$, the distance (henceforth depth measurement) between the agent's location and the point on the obstacle's boundary curve intersecting the depth sensor vector (henceforth impact point) and (ii) $\psi$, the angular separation between $\dot{f}$ at the impact point and the depth sensor vector. If the depth measurement is not less than $d_m$, the agent does not detect the presence of obstacles and the sensor returns only the depth measurement, i.e. $d=d_m$. Control of the sensor orientation is modelled in continuous time,
\begin{equation}
    \dot{\phi}=u_s,
\end{equation}
where the sensor is initially parallel to the velocity, i.e.  $\phi(0)=0$, and the sensor actuation is limited $\left \lvert u_s \right \rvert \leq \gamma=\frac{1}{10}\frac{a_0}{v_0}$.\\
The goal of this case case study is to design for $u$ and $u_s$ and theoretically guarantee that the agent reaches the origin while avoiding collisions with obstacles.
 \section{Control Algorithm}
In this section we describe the control algorithm that will certify convergence to the origin of the agent. The idea is to track the line passing through the origin with gradient $v(0)$ (henceforth termed the attraction line) in absence of obstacles in the agent's local horizon. Once an obstacle is detected, the agent should follow the obstacle at a sufficiently close distance so that no other obstacles appear in the agent's horizon (this phase will be called the tracking). The obstacle should stop following the obstacle once it is in in proximity of the attraction line. Note that, once an obstacle is detected the agent needs to reorient its velocity and sensor direction (recall that in absence of obstacles the sensor is parallel to the velocity) in order to be able to handle tracking, i.e. during tracking we expect the velocity to be almost parallel to the tangent of the obstacle curve and the sensor to be orthogonal to the velocity. The latter phase of reorientation will be termed locking. Similarly, the agent needs to adjust its orientation once tracking ends in favour of following the attraction line. This phase is called unlocking. We will thus dedicate the reminder of this article in finding an appropriate control law for tracking, locking and unlocking. Note that a controller has to also be designed in order to track the attraction line, but as is described in \S\ref{sec:locking}, we can use the control designed for the immediate locking section.
\subsection{Controller Modes}\label{sec:control}
The controller has four modes corresponding to (i) following the attraction line, (ii) locking, (iii) tracking and (iv) unlocking.\\
Mode (i) defines the controller with which the algorithm starts and is the mode used in absence of obstacles in the local horizon of the agent (i.e. when $d=d_m$). Note that since no obstacles are present, $u_s=0$. This mode ensures that the agent follows the attraction line, by ensuring that (i) the agent's location, $x$, is in a tubular neighbourhood of radius $\Delta_m r_0=0.03r_0$ of the attraction line and (ii) the angular separation between $v$ and $v(0)$ (denoted by $\delta$) is not greater than $\delta_m=0.103$. The controller for this mode is derived in \S\ref{sec:locking} and is given by,
\begin{equation}
    u_{att}\left(x,v,x(0),v(0)\right)=-\delta-163\Delta,
    \label{eq:u_att}
\end{equation}
where $\Delta=\pi_{att}\left(x\right)$ and $\pi_{att}:\mathbb{R}^2\rightarrow\mathbb{R}_{\geq 0}$ represents distance between the agent's position and its projection onto the attraction line. Mode switching occurs if either (a) $x\in B\left(0,\epsilon\right)$ where $\epsilon\leq d_m$ at which point the algorithm stops or (b) $d<d_m$ at which point Mode (ii) begins and a timer, with values $\hat{t}$, begins.\\
For locking, the depth sensor detects the initial impact point on the obstacle (the first point at which $d<d_m$) and keeps tracking it at all times (see \S\ref{sec:locking} for a mathematical expression for $u_s$). Using the latter point, it is possible to obtain nominal trajectories, $(\overline{x},\overline{v})$, which end when $\overline{v}$ is aligned with $\dot{f}$ at the impact point and the distance between $\overline{x}$ and the impact point is $d_0=0.06r_0$ (the time of completion for the nominal trajectory is given by $\hat{t}_1$). The proposed nominal trajectory involves three segments (and thus two switching submodes): (a) an initial circular segment (with radius $1.151r_a$), (b) an intermediate linear segment and (c) a final circular segment (with radius $1.151r_a$). A controller is designed for each submode which ensures that the agent tracks the nominal trajectory (analogous to mode (i)). For the initial and final submodes, the controller is given by,
\begin{equation*}
    u_{cir}\left(x,v,\overline{x},\overline{v},\hat{t}\right)= u=-M\text{sgn}\left(\delta-3.4\Delta\right),
    \label{eq:circ}
\end{equation*}
where $\delta$ and $\Delta$ have analogous meanings to the ones for mode (i) (e.g. $\Delta$ is the distance between the agent's location and its projection on the nominal curve, etc). Similarly, since the intermediate stage is linear we define $u_{int}$ as in (\ref{eq:u_att}) with obvious modifications (see \S\ref{sec:locking} for details). Note that for each submode, trajectories are guaranteed to exits either by the continuity of the controller (intermediate stage) or by convexification\cite{Filippov} (initial and final stages) and hence since there are a finite number of switches (two) between the submodes, a trajectory exists for the entire locking stage. We will refer to $u_{loc}$ as the collection of the three submodes that produces the aforementioned trajectory. Switching from mode (ii) to mode (iii) occurs when $\hat{t}\geq \hat{t}_1$, the distance between $x$ and the impact point is contained in $\left[\frac{d_0}{2},\frac{3d_0}{2}\right]$ and the angular separation between $v$ and $\dot{f}$ at the impact point is bounded by $\psi_m=0.58$. At the time of the switch the timer is reset.\\ 
We define $\hat{t}_2$ to be the time taken to reach the attraction line by a particle travelling in uniform circular motion with acceleration $a_0$ and starting position and velocity equal to agent's when the mode switching occurred. In mode (iii), the controller is given by,
\begin{equation*}
    u_{tr}\left(x,v,d,\psi,\hat{t}\right)=-M\text{sgn}\left[0.02\left(d-d_0\right)-0.12\left(\psi-1.71\right)\right]
\end{equation*}
and guarantees that $d \in \left[\frac{d_0}{2},\frac{3d_0}{2}\right]$ and $\lvert \psi \rvert \leq \psi_m$ at all times. Throughout tracking, $u_s=0$. Switching from mode (iii) to mode (iv) occurs when $\hat{t}\geq \hat{t}_2$ and $x$ is the neighbourhood of radius $\Delta_mr_0$ of the attraction line. Mode switching resets the timer for $\hat{t}$.\\
Mode (iv) is analogous to mode (ii) in the sense that the sensor is oriented towards a fixed impact point (the one detected when mode switching from (iii) to (iv) occurs). Similarly, nominal trajectories involving only two circular segments can be used to drive the state of the agent to have velocity parallel to $v(0)$ and position on the attraction line and $\hat{t}_3$ is defined analogously. Note that at this point, the sensor will not be parallel to the velocity so it has to be rotated using maximal sensor actuation until $\phi=0$ (details in \S\ref{sec:unlocking}). The circular submodes are defined just as in mode (ii) and $u_{unloc}$ is used to denote the collection of the two circlar submodes and submode of sensor reorientation. Mode (iv) switches to mode (i) when $\hat{t}\geq \hat{t}_3$, the angular separation between $v$ and $v(0)$ is bounded by $\delta_m$ and $x$ is the tubular neighbourhood of radius $\Delta_mr_0$ of the attraction line. A summary of how mode switching occurs is presented in Algorithm 1.\\
It is important to emphasize that the controller modes designed guarantee sensitivity to some level of sensor noise (see \S\ref{sec:tracking} for example). However, no theoretical guarantees are given on the magnitude of the noise, i.e. for tracking there exists $\delta_{tr}$ such that if sensor noise is bounded by $\delta_{tr}$, tracking is guaranteed to be possible. $\delta_{loc}$ and $\delta_{unloc}$ are similarly defined. Moreover, we stress that given the assumptions on the family of obstacles and the agent dynamics, it is possible to infer that each mode is guaranteed to occur for a finite non-infinitesimal amount of time after every mode switch. This implies that convergence to the target will only require a finite amount of mode switches and since each mode guarantees the existence of a trajectory associated with it, wellposedness of (\ref{eq:agent_dynamics}) follows.
\section{Main Result}
We now state the main theorem of this article that rigorously proves that the aforementioned control algorithm will guarantee, that if the agent starts sufficiently far from an obstacle, it will reach a neighbourhood of the target while avoiding collisions with obstacles at all times.
\begin{theorem}
Assume there are no obstacles in $B(0,d_m)$ and $B(x(0),d_m)$. Then for every $\epsilon>0$, there exists $\delta>0$ and finite time $T>0$ such that for sensor noise bounded by $\delta$ and for any $x(0)$, we have $x(T) \in B(0,\epsilon)$.
\end{theorem}
\noindent Proving Theorem 1 is straightforward since tracking, locking and unlocking have already been certified. Consider the case of zero noise, then, there exists a continuous trajectory that always ends on the attraction line after unlocking and hence there exists a finite time $T$ such that $x(T)=0$. Similarly, we set $\delta=\min\left(\delta_{tr},\delta_{loc},\delta_{unloc}\right)$ and we introduce noise bounded by $\delta$ in sensor measurements. By the continuity of the trajectory, noise in sensor measurements implies that after unlocking the agent's location and velocity are in a neighbourhood of the attraction line and $v(0)$ respectively, whence proving the desired result. \\
Note that it is important to emphasize that Theorem 1 proves robustness to some non-zero level of noise, however we do not actually prove a global upper bound for the noise.
\section{Numerical Examples}\label{sec:simulations}
We conclude our case study by providing some simulations of the agent moving in an unknown environment. We keep the sensor noise at $\frac{d_0}{8}$ for depth measurements and $1.5\degree$ for angular measurements and we stress that the latter are not theoretically guaranteed but rather the practical (obtained from numerical experiments) maximal noise levels that the proposed algorithm can handle. Figure 1 (a) depicts the agent moving past an unknown of family of obstacles and converging to the target. On the other hand, in Figure 1 (b) we portray the importance of condition (ii) in Assumption 1 for the locking phase by showing fatal collision in absence of the latter condition. Finally, in Figure 1 (c)-(d) we compare tracking for bounded obstacle curvature (Figure 1 (c)) versus the inability to track obstacles with large curvature (Figure 1 (d)).
\begin{figure}[H]
\centering
\begin{subfigure}[b]{0.48\textwidth}
\subcaption{}
\centering
\includegraphics[scale=0.48,trim={1cm 0.4cm 1.2cm 0cm},clip]{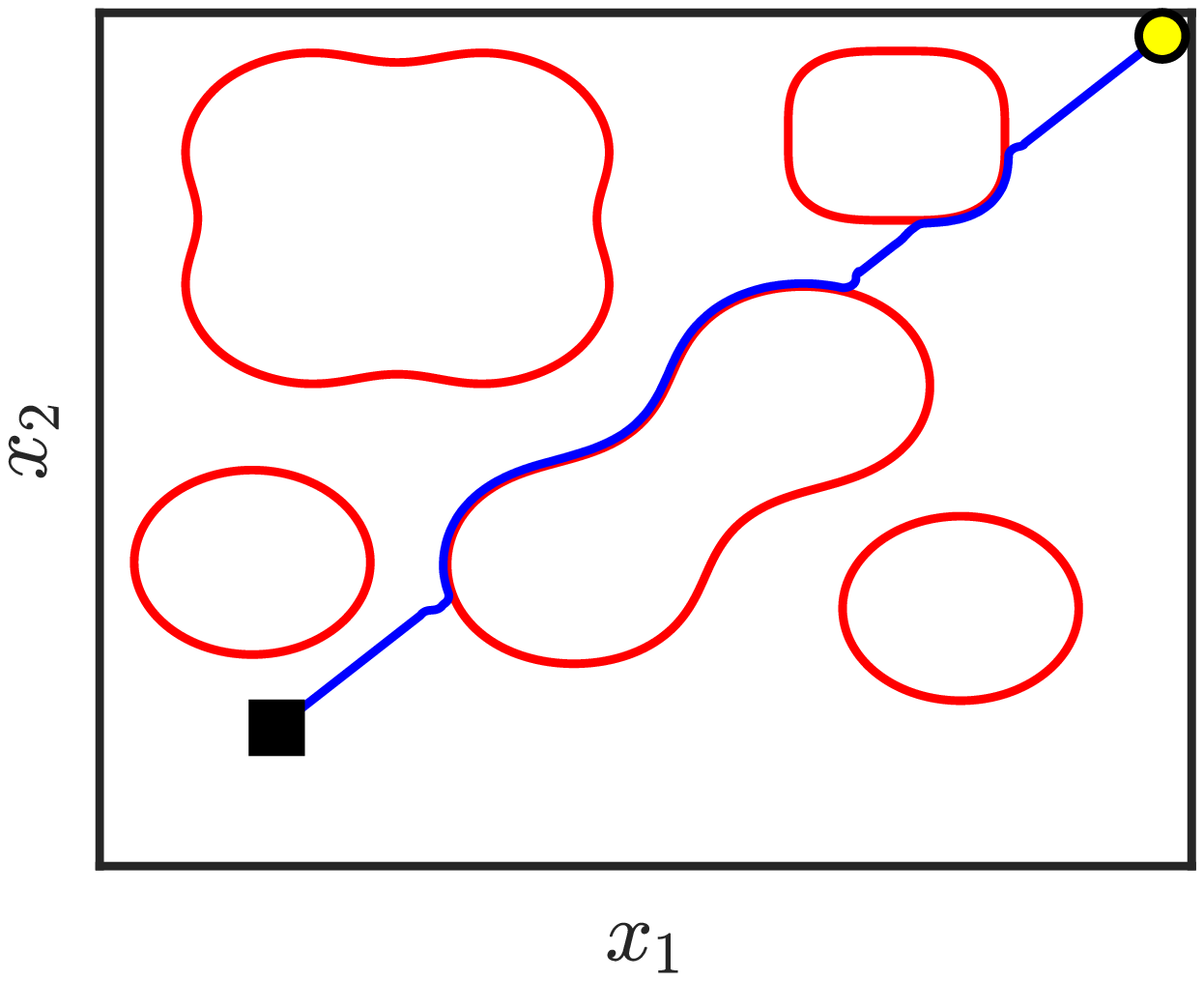}\label{figa}
\end{subfigure}
\hfill
\begin{subfigure}[b]{0.48\textwidth}
\subcaption{}
\centering
\includegraphics[scale=0.48,trim={1cm 0.4cm 1.2cm 0cm},clip]{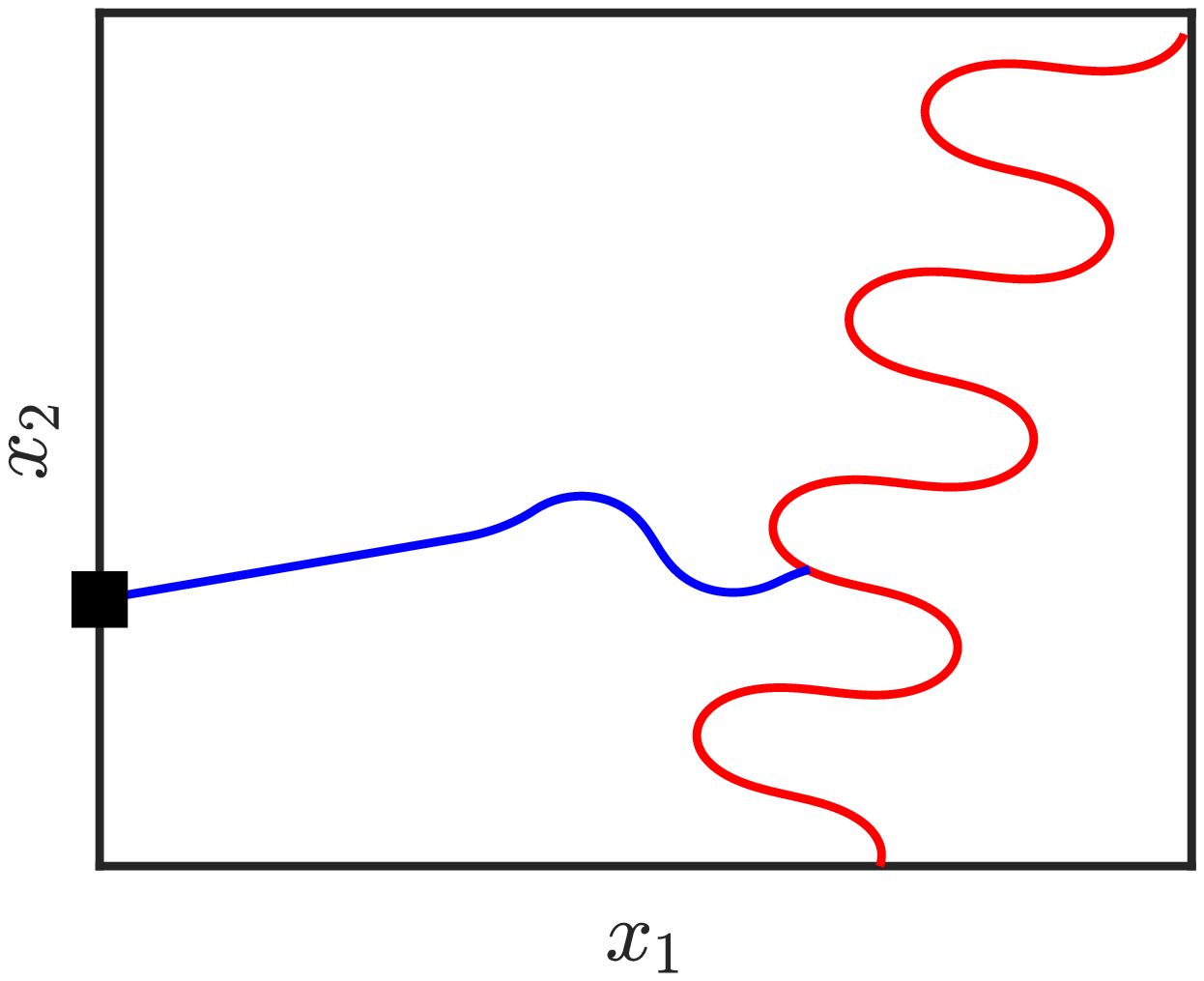}\label{fig:b}
\end{subfigure}
\begin{subfigure}[b]{0.48\textwidth}
\subcaption{}
\centering
\includegraphics[scale=0.48,trim={1cm 0.2cm 1.2cm 0cm},clip]{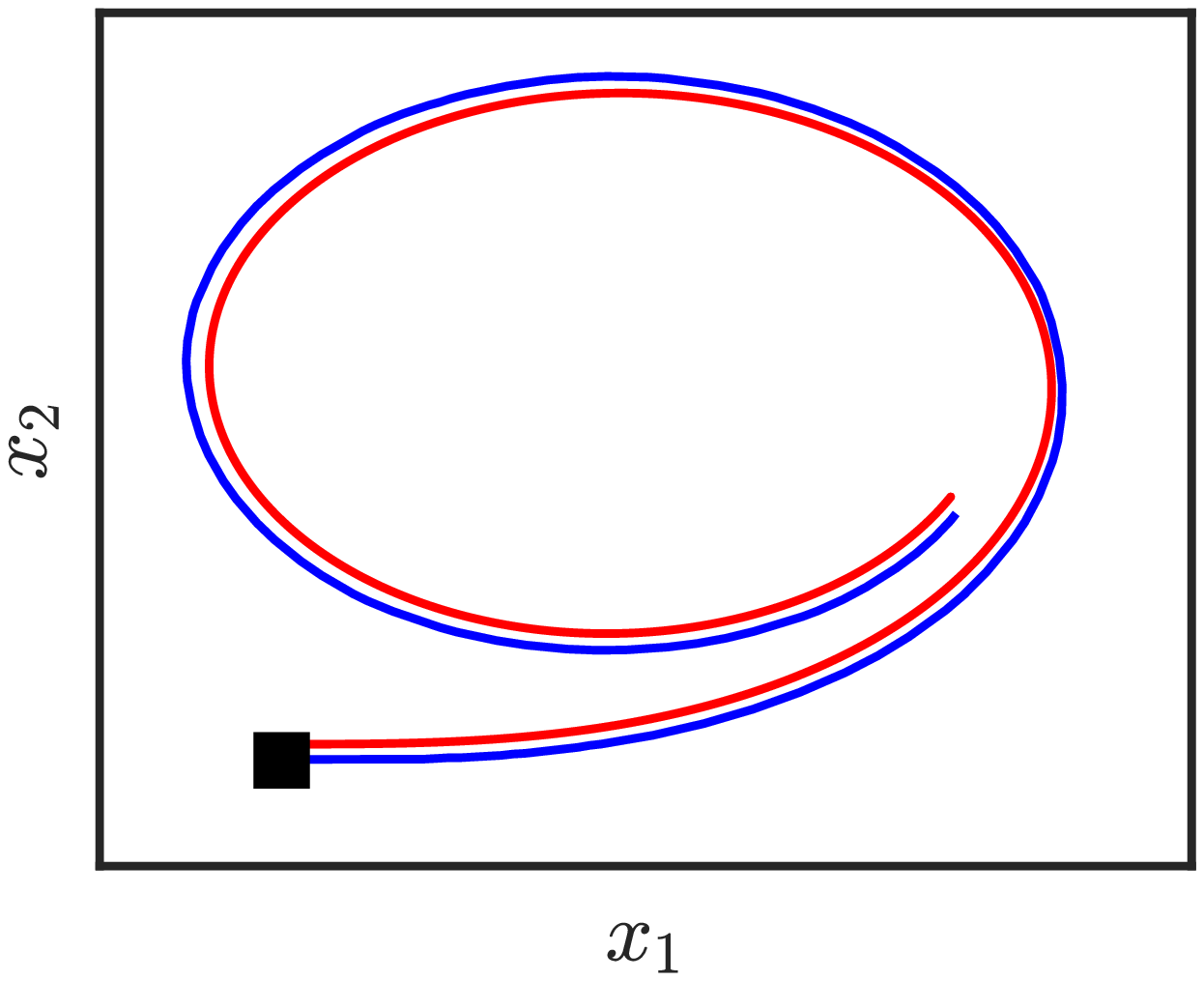}\label{fig:c}
\end{subfigure}
\hfill
\begin{subfigure}[b]{0.48\textwidth}
\subcaption{}
\centering
\includegraphics[scale=0.48,trim={1cm 0.2cm 1.2cm 0cm},clip]{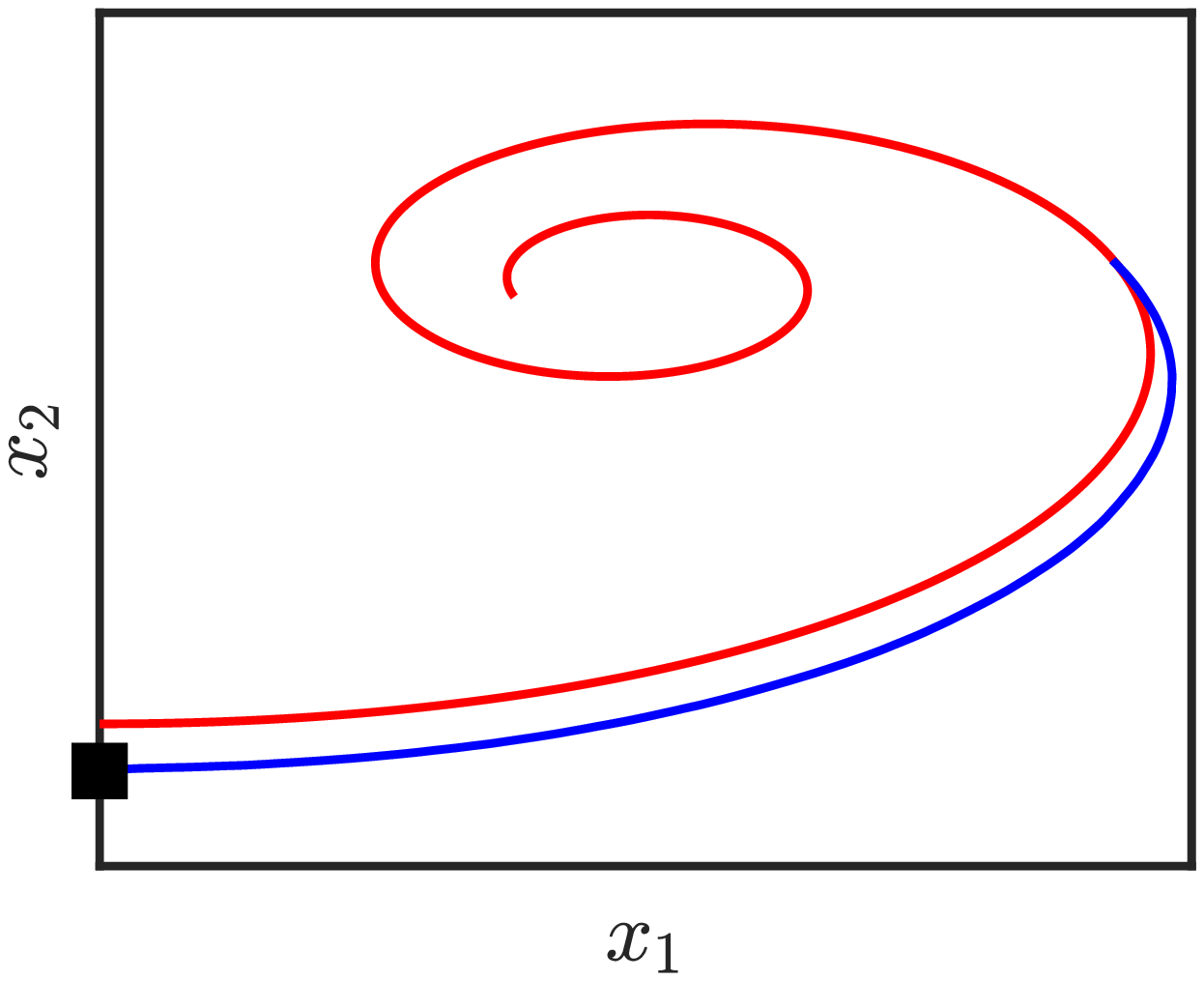}\label{fig:d}
\end{subfigure}
  \caption{(a) Convergence to target, (b) dropping condition (ii) from Assumption 1. Importance of curvature assumption for tracking: (c) tracking with $w(s)=\frac{2}{\pi}\arctan\left(s\right)$ vs (d) tracking with $w(s)={\rm e}^s$. Obstacles portrayed in red, starting position of agent is denoted by $\blacksquare$ and the target is denoted by $\bullet$.}\label{fig1}
\end{figure}
\bibliography{references}
\appendix
\section{Derivations}
\subsection{Tracking}\label{sec:tracking}
We derive the control law for tracking a single obstacle and hence we drop the obstacle subscript $i$. Let $T$ denote the rotation matrix by $\phi$ where $\phi$ is the angle between the velocity and depth sensor and define the angle between the depth sensor and the tangent to the obstacle curve as $\psi$, then
\begin{equation}
    v_0^2\cos(\phi-\psi)=v'\dot{f}.
    \label{eq:angledef}
\end{equation}
The dynamics of the agent, sensor depth, $d$, and obstacle boundary are related by,
\begin{equation}
    x+d\frac{Tv}{v_0}=f,
    \label{eq:depthsenor}
\end{equation}
so that by differentiating (\ref{eq:depthsenor}) and combining with (\ref{eq:angledef}), we can obtain dynamic equations for $d$ and $\psi$. Henceforth, in this section, we work with non-dimensional depth $d \rightarrow \frac{d}{r_0}$ and time, $t \rightarrow \frac{a_0}{v_0}t$ and obtain,
\begin{eqnarray}
    \dot{d}&=&-\cos{\phi}+\frac{\sin{\phi-du}}{\tan{\psi}}, \nonumber \\ 
    \dot{\phi}&=&0, \nonumber \\
    \dot{\psi}&=&u-\frac{\sin{\phi-du}}{\sin{\psi}}w.
\end{eqnarray}
For the tracking phase, the idea is to keep the agent within a neighbourhood of the obstacle curve and have the velocity track the obstacle tangent as close as possible. Thus, the aim is to keep $d \in [\frac{d_0}{2},\frac{3d_0}{2}]$ and $\psi\in[\frac{\pi}{2}-\psi_m,\frac{\pi}{2}+\psi_m]$ where $0<d_0\ll 1$ and $\psi_m>0$. Writing $\Delta=d-d_0$, $\phi':=\phi-\frac{\pi}{2}$ and $\delta:=\psi-\frac{\pi}{2}$, then, $\dot{\phi'}=0 \implies \phi':=\phi_m\geq0$ (we drop this non-negativity assumption at the end of this section) and 
\begin{eqnarray}
    \dot{\Delta}&=&\sin{\phi_m}-(\cos{\phi_m}-(\Delta+d_0)u)\tan{\delta}, \nonumber \\ 
    \dot{\delta}&=&u-\left(\frac{\cos{\phi_m}-(\Delta+d_0)u}{\cos{\delta}}\right)w. 
\end{eqnarray}
In order, to obtain a control law for the above, we consider the inner approximation
\begin{eqnarray}
    \dot{\Delta}&=&\sin\phi_m-\alpha \delta, \nonumber \\
    \dot{\delta}&=&u-\beta,
\end{eqnarray}
where
\begin{eqnarray*}
    \alpha &\in& [\alpha_0,\alpha_1]= \left[\frac{2\cos{\phi_m}-3d_0M}{2},\frac{2\cos{\phi_m}+3d_0M}{2}\right], \\
    \beta &\in& [-\beta_0,\beta_0]= \left[-\frac{2\cos{\phi_m}+3d_0M}{2\cos{\psi_m}},\frac{2\cos{\phi_m}+3d_0M}{2\cos{\psi_m}}\right].
\end{eqnarray*}
This immediately yields some conditions for stabilizability,
\begin{eqnarray*}
    d_0M<\frac{2}{3}\cos{\phi_m}, \hspace{0.5cm} \text{and} \hspace{0.5cm} M>\frac{2\cos{\phi_m}+3d_0M}{2\cos{\psi_m}}.
\end{eqnarray*}
Let $\psi_0:=\frac{\sin\phi_m}{2}\left(\frac{1}{a_0}+\frac{1}{a_1}\right)$ and impose $\psi_m>\frac{sin(\phi_m)}{a_0}$ and $\frac{\sin{\phi_m}}{a_1}<\phi_m<\frac{\sin{\phi_m}}{a_0}$. We define a quadratic control Lyapunov function,
\begin{equation}
    V=\Delta^2+p^2(\delta-\psi_0)^2+2q\Delta(\delta-\psi_0),
\end{equation}
and seek the region of attraction for the tracking phase by imposing $\dot{V}<0$,
\begin{equation*}
    (\Delta+q\left(\delta-\psi_0\right))\left(\sin\phi_m-\alpha\delta\right)-\beta(p^2\Delta+q\left(\delta-\psi_0\right))<M\left\lvert p^2\Delta+q\left(\delta-\psi_0\right)\right\rvert,
    \label{eq:trackingclf}
\end{equation*}
where we set $u=-M\text{sgn}\left(p^2\Delta+q\left(\delta-\psi_0\right)\right)$. For convenience, we find a subset for the region described above,
\begin{eqnarray}
        \hat{m}\left(\delta+\hat{c}\right)\leq \Delta \leq \hat{m}\left(\delta-\hat{c}\right),
 \label{eq:boundary}
\end{eqnarray}
where $\hat{m}=\frac{q}{p^2-q^2}\frac{q\sin{\phi_m}-(M-\beta_1)p^2-\alpha_1q\psi_0}{M-\beta_1}$ and $\hat{c}=-\frac{\sin{\phi_m}}{\alpha_1}+\frac{(M-\beta_1)p^2}{\alpha_1q}$. However, considering this simplified region yields yet another stability condition,
\begin{equation*}
    \frac{\sin{\phi_m}}{\alpha_1}-\frac{p^2}{\alpha_1q}(M-\beta_1)>\psi_0.
\end{equation*}
Before proceeding, we summarise all constraints:
 \begin{subequations}
 \begin{eqnarray}
     \psi_m&>&\frac{\sin{\phi_m}}{\alpha_0}, \label{eq:condM1}\\
     \phi_m&>&\frac{\sin{\phi_m}}{\alpha_1} \label{eq:condM2}\\
     2\cos{\phi_m}-3d_0M&>&0,\label{eq:d0}\\
     2M\cos\psi_m&>&2\cos{\phi_m}-3d_0M,\label{eq:psi}\\
      \frac{\sin{\phi_m}}{\alpha_1}-\frac{p^2}{\alpha_1q}(M-\beta_1)&>&\psi_0\label{eq:condp}.
 \end{eqnarray}
 \end{subequations}
We wish to maximize the region of attraction for tracking (remember that we need to bound the control Lyapunov function ellipse so that it is physically relevant, i.e. $\lvert \Delta \rvert \leq\frac{d_0}{2}$ and $\lvert \delta \rvert \leq \psi_m-\psi_0$). Whence to satisfy (\ref{eq:d0}) and (\ref{eq:psi}), we select,
\begin{equation*}
    3d_0M=\cos{\phi_m} \hspace{0.5cm} \text{and} \hspace{0.5cm} \cos{\psi_m}=\frac{2\cos{\phi_m}}{M}.
\end{equation*}
This ensures that $d_0$ and $\psi_m$ are maximised and also simplifies the parameters,
$\alpha_0=\frac{\cos\phi_m}{2}$, $\alpha_1=\frac{3\cos\phi_m}{2}$, $\beta_1=\frac{3M}{4}$ $\psi_0=\frac{4\tan\phi_m}{3}$. Moreover, we impose $M\geq3$ and $\phi_m <0.431$ in order to satisfy (\ref{eq:condM1}) and (\ref{eq:condM2}).\\
In order to find the largest region of attraction we impose that the elliptic control Lyapunov at $(-\psi_0+\frac{\sin{\phi_m}}{\alpha_1}-\frac{p^2}{\alpha_1q}(M-\beta_1),0)$  has the same tangent gradient as the two boundary lines in (\ref{eq:boundary}). 
This is achieved by selecting,
\begin{equation}
    p^2=q^2-q\sqrt{q^2+\frac{4q}{M}\sin{\phi_m}}.
    \label{eq:p}
\end{equation}
Note that the condition in (\ref{eq:condp}) is equivalent to imposing $q\leq -\frac{4}{M}\sin{\phi_m}$.\\
We proceed by fitting the ellipse inside the correct box in order to be physically relevant and thus require,
\begin{equation*}
    \frac{p^2\left(\frac{2\tan{\phi_m}}{3}+\frac{Mp^2}{6q\cos{\phi_m}}\right)^2}{p^2-q^2}\leq \frac{\cos^2{\phi_m}}{36M^2}\leq \left(\psi_m-\psi_0\right)^2,
\end{equation*}
which is equivalent to requiring that $(\phi_m,q)$ is in the intersection:
\begin{equation*}
    \left\{0>q\geq -\frac{1}{8^{\frac{1}{4}}M}\right\}\bigcap\left\{q\leq \left(\frac{0.109}{\sqrt{\arccos{\frac{2}{M}}}}-0.775\right)\phi_m-\frac{0.236}{M\sqrt{\arccos{\frac{2}{M}}}}\right\}.
\end{equation*}
Thus, to maximize $\phi_m$, we select, $q= -\frac{1}{8^{\frac{1}{4}}M}$ and obtain
\begin{equation*}
    \phi_m \leq \frac{\left(\frac{1}{3M\sqrt{2\arccos\left(\frac{2}{M}\right)}}-\frac{1}{8^{\frac{1}{4}}M}\right)\left(1.293M\sqrt{2\arccos\left(\frac{2}{M}\right)}\right)}{\left(1-5.013\sqrt{2\arccos\left(\frac{2}{M}\right)}\right)}.
\end{equation*}
We conclude the tracking section by bounding the spurious region around the origin where $\dot{V}\geq 0$. The latter is contained inside the following ellipse,
\begin{equation}
    \Delta^2+p^2\left(\delta-\psi_0\right)^2+2q\Delta\left(\delta-\psi_0\right)\leq \frac{4p^2}{9q^2}\left(p^2-q^2\right)\tan^2\phi_m,
    \label{eq:spurios}
\end{equation}
whereas the maximal region of attraction is given by the intersection of the complement of the region in (\ref{eq:spurios}) with the region below,
\begin{equation}
    \Delta^2+p^2\left(\delta-\psi_0\right)^2+2q\Delta\left(\delta-\psi_0\right)\leq p^2\left(\frac{2}{3}\tan\phi_m+\frac{Mp^2}{6q\cos\phi_m}\right)^2.
\end{equation}
We also stress that by symmetry, the condition $\phi_m\geq0$ can be dropped but (\ref{eq:p}) is replaced with
\begin{equation}
     p^2=q^2-q\sqrt{q^2+\frac{4q}{M}\sin{\phi_m}\text{sgn}(\phi_m)},
\end{equation}
which ensures that the rest of the analysis remains valid. \\
Hence, thus far we have proved the existence of $\delta_{tr}$ such that if the sensor noise is bounded by $\delta_{tr}$ then the agent is guaranteed to be able to track an obstacle (this follows from the fact that the Lyapunov function is strictly decreasing). Finally, note that while $M \geq 3$, we arbitrarily select $M=5$ in order to ensure that tracking occurs close enough to the obstacle (i.e $d_0\approx\frac{6}{100}r_0$) and a decent sensitivity to sensor noise (empirically determined via simulations, see \S\ref{sec:simulations}).
\subsection{Locking}\label{sec:locking}
Consider the moment when the agent first detects the presence of the obstacle. At this time instance, the coordinates of the impact point, $f_{p}$, and the tangent at the later point, $\dot{f}_p$, will be known (subject to noise). In order to simplify the analysis in the following section, we define coordinates in the complex plane, $\mathbb{C}$, and tangents/velocities on the unit circle, $\mathbb{T}$, with the understanding that if $x \in \mathbb{C}$ then $\text{Re}(x)$ represents the horizontal coordinate, $\text{Im}(x)$ represent the vertical coordinate, etc. We, thus define local coordinates with the local origin at $f_{p}$ and define the real axis to be parallel to $\dot{f}_p$ while the imaginary axis is parallel to the outward normal to the obstacle curve at $f_p$. Note that by defining such a coordinate axis we obtain a control law in local coordinates which will then need to be converted in the global coordinate frame.\\
The objective is to drive the agent to the location $jd_0$ with velocity $v_0$ where $j=\sqrt{-1}$ and $v_0>0$ is the magnitude of the constant velocity of the agent. Note that we do this by keeping the impact point of the sensor at $f_p$. In the absence of noise, it is straightforward to define a nominal trajectory bringing the agent to the desired state. Namely,
\begin{subequations}
\begin{align}
\begin{split}
\begin{cases}
\overline{x}(t)=\left(-d_m+j\frac{r_0}{\overline{u}}\right){\rm e}^{-j\theta}+\frac{r_0}{\overline{u}}{\rm e}^{j\left(\frac{a_0\overline{u}}{v_0}t-\theta-\frac{\pi}{2}\right)}\\
\overline{v}(t)=v_0{\rm e}^{j\left(\frac{a_0\overline{u}}{v_0}t-\theta\right)}\\
\overline{u}(t)=\frac{M}{c_1(\theta)\left(1+\epsilon\right)}
\end{cases}
t\in [0,t_1]
\end{split}\\
\begin{split}
\begin{cases}
\overline{x}(t)=\overline{x}(t_1)+\overline{v}(t_1)(t-t_1)\\
\overline{v}(t)=\overline{v}(t_1)\\
\overline{u}(t)=0
\end{cases}
\hspace{2cm}t\in (t_1,t_2]
\end{split}\\
\begin{split}
\begin{cases}
\overline{x}(t)=j\left(d_0+\frac{r_0}{\overline{u}}\right)+\frac{r_0}{\overline{u}}{\rm e}^{j\left(\frac{a_0\overline{u}}{v_0}(t-T)-\frac{\pi}{2}\right)}\\
\overline{v}(t)=v_0{\rm e}^{j\left(\frac{a_0\overline{u}}{v_0}(t-T)\right)}\\
\overline{u}(t)=\frac{M}{c_2(\theta)\left(1+\epsilon\right)}
\end{cases}
\hspace{0.5cm}t\in (t_2,T]
\end{split}
\end{align}
\end{subequations}
where we recall that $d_m$ is the maximal depth of the sensor, $\cos{\theta}:=\frac{\text{Re}(\overline{v}(0))}{v_0}$, $d_0$ is the desired distance from the obstacle required for tracking, $r_0$ is the minimum radius of curvature of the obstacle boundary curve, $a_0M$ is the maximum allowable actuation (obtained from tracking), $t_1,t_2,T$ ensure that the nominal trajectory is continuous and that the intermediate linear trajectory (i.e. $t \in [t_1,t_2])$ is the common tangent connecting the initial (i.e. $t \in [0,t_1]$) and final (i.e. $t \in [t_2,T]$) circular segments.  $\epsilon>0$ parametrizes the increase in radius, with respect to the minimum radius of the agent, necessary in order to handle noise and will be calculated when noise is introduced. Moreover, for future convenience we define, $\hat{x}_1(\overline{u}):=\left(-d_m+j\frac{r_0}{\overline{u}}\right){\rm e}^{j\theta}$, $\hat{x}_2:=\overline{x}(t_1)$ and
$\hat{x}_3(\overline{u}):=j\left(d_0-\frac{r_0}{\overline{u}}\right)$, $\hat{x}_{1,3}^+:=\hat{x}_{1,3}\left(\lvert \overline{u} \rvert\right)$, $c_2(\theta):=\text{sgn}\left(a(\theta)\right)$ where 
\begin{equation*}
    a(\theta):=\frac{2r_0(1+\epsilon)}{M}\text{Re}(\hat{x}_1^+)+\text{Im}(\hat{x}_1^+-\hat{x}_3^+)\sqrt{\text{Re}(\hat{x_1}^+)^2+\text{Im}(\hat{x}_1^+-\hat{x}_3^+)^2-\frac{4r_0^2(1+\epsilon)^2}{M^2}}.
\end{equation*}
Similarly, $c_1(\theta):=\text{sgn}\left(\cos{\theta}\left(1+\frac{Md_0}{r_0(1+\epsilon)}\right)-1\right)$.\\
We then proceed with the introduction of disturbances of the appropriate form in the nominal trajectories. For the initial ($t\in [0,t_1]$) and final ($t \in (t_2,T]$) stages, we write, 
\begin{subequations}
\begin{equation}
    \begin{cases}
    x(t)=\hat{x}_{1,3}+r_0\left[\frac{1}{\overline{u}}+\Delta(t)\right]{\rm e}^{j\overline{u}q(t)},\\
    v(t)=jv_0{\rm e}^{j\left(\overline{u}q(t)+\delta(t)\right)},
    \end{cases}
\end{equation}
where $\delta,\Delta, q \in \mathbb{R}$, are non-dimensional and satisfy
\begin{equation}
    \begin{cases}
    \tan\delta:=-\frac{\text{Re}\left(\left(x-\hat{x}_{1,3}\right)'v\right)}{\text{Im}\left(\left(x-\hat{x}_{1,3}\right)'v\right)},\\
    \Delta:=\frac{\left\lvert x-\hat{x}_{1,3}\right\rvert}{r_0} -\frac{1}{\overline{u}}.
\end{cases}
\end{equation}
\end{subequations}
Similarly for the intermediate stage ($t\in(t_1,t_2]$),
\begin{subequations}
\begin{equation}
    \begin{cases}
    x(t)=\hat{x}_{2}+\frac{v_0\overline{v}(t)}{a_0}\left[q(t)+j\Delta\right],\\
    v(t)=\overline{v}(t){\rm e}^{j\delta(t)},
    \end{cases}
\end{equation}
with,
\begin{equation}
    \begin{cases}
    \tan\delta:=-\frac{\text{Im}\left(v'\overline{v}\right)}{\text{Re}\left(v'\overline{v}\right)},\\
    \Delta:=-\frac{1}{v_0r_0}\text{Im}\left(\left(x-\hat{x}_{2}\right)'\overline{v}\right).
\end{cases}
\end{equation}
\end{subequations}
Moreover, we can derive the dynamic equations for the disturbances, in the similar way as we did for the tracking case. We recall that we defined the non-dimensional the time, $\tilde{t}=\frac{a_0}{v_0}t$ whence the dynamic non-dimensional in time equations for the initial and final stage are given by
\begin{subequations}
\begin{align}
\dot{\delta}&=u-\overline{u}\dot{q}, \nonumber \\
\dot{\Delta}&=-\sin{\delta}, \nonumber \\
\dot{q}&=\frac{\cos{\delta}}{1+\overline{u}\Delta}, \label{eqcontrol:circle}
\end{align}
whereas for the intermediate stage,
\begin{align}
\dot{\delta}&=u-\overline{u}\dot{q}, \nonumber \\
\dot{\Delta}&=-\sin{\delta}, \nonumber \\
\dot{q}&=\frac{\cos{\delta}}{1+\overline{u}\Delta}. \label{eqcontrol:linear}
\end{align}
\end{subequations}
We stress here that the necessary space required for locking is given by the tubular neighbourhood of the obstacle that has radius $r_0\left(\frac{3\left(1+\epsilon\right)}{M}+\frac{3d_0}{2}\right)<r_0$. This thus imposes the minimal allowable sensor depth. Note that the aforementioned distance was obtained by considering the agent approaching a convex obstacle with velocity orthogonal to obstacle tangent. The above argument also remains valid for non-convex obstacles since by assumption at each point on the obstacle boundary curve there are no other boundary curves within a ball of radius $r_0$ (condition (ii) in Theorem 1).\\
Furthermore, we can compute the minimal actuation rate of the sensor, $\gamma$, by analysing the dynamic equations for $d,\phi$ and $\psi$ in the locking phase (derivation analogous to tracking),
\begin{eqnarray}
    \dot{d}&=&-v_0\cos{\phi}, \nonumber \\ 
    \dot{\phi}&=&\frac{v_0\sin{\phi}}{d}-\frac{a_0}{v_0}u \nonumber \\
    \dot{\psi}&=&\frac{v_0\sin{\phi}}{d},
\end{eqnarray}
which thus implies $\gamma\geq\frac{a_0}{v_0}\left(\frac{M}{1+\epsilon}+\frac{2r_0}{d_0}\right)$.\\
We conclude this by designing the control law which proves that locking is possible under our assumptions. We begin with the initial and final circular motion segments. We wish to keep the disturbances in $\lvert \delta \rvert \leq \delta_m \leq \psi_m-\phi_m $, $\lvert \Delta \rvert \leq \Delta_m \leq \frac{d_0}{2}$ (so that noise does not interfere with the agent's ability to track) and $\dot{q} \approx 1$. Note that from equation (\ref{eqcontrol:circle}) it suffices to keep $\delta, \Delta$ small in order to maintain  $\dot{q} \approx 1$. We proceed by obtaining an inner approximation for the equation (\ref{eqcontrol:circle}), 
\begin{align}
\dot{\delta}&=u-\beta, \nonumber \\
\dot{\Delta}&=-\alpha\delta,
\end{align}
where
\begin{equation*}
    \alpha \in \left[\frac{\sin{\delta_m}}{\delta_m},1\right] \,\,\, \text{and} \,\,\, \beta \in \left[-\frac{\lvert \overline{u}\rvert}{1-\Delta_m\lvert \overline{u}\rvert},\frac{\lvert \overline{u}\rvert}{1-\Delta_m\lvert \overline{u}\rvert}\right].
\end{equation*}
and we note that we have reduced this problem to the one solved in the tracking case. So omitting most details, our control is defined as 
\begin{equation}
    u=-M\text{sgn}\left(\delta-\frac{p}{\sqrt{2}}\Delta\right)
\end{equation}
where $p=\frac{2M}{\sqrt{2+\epsilon}}$ and the control Lyapunov function is defined as the quadratic,
\begin{equation}
    V:=\delta^2+p^2\Delta-\sqrt{2}p\delta\Delta.
\end{equation}
 Maximimizing the region of attraction for the above control Lyapunov function we obtain the bounds, $\Delta_m \leq \frac{\epsilon}{2M}$ and $\delta_m\leq \frac{\epsilon}{\sqrt{2+\epsilon}}$. Whence in order to comply with the tracking requirement we require $\epsilon \leq 0.151$. We maximise $\epsilon$ in order to maximize the noise that can be handled and hence we obtain $\delta_m = 0.103$ and $\Delta_m = 0.030$, $p=6.818$.\\
Controlling the linear intermediate portion is simpler and does not require discontinuous feedback. The requirement that $\dot{q} \approx 1$ in equation (\ref{eqcontrol:linear}) is analogous to initial/final cases we just studied and thus only requires that $\lvert \delta \rvert \leq \delta_m$, $\lvert \Delta \rvert \leq \Delta_m$. We define the control Lyapunov function as
\begin{equation}
    V=\delta^2+q^2\Delta^2,
\end{equation}
and the control law as $u=-\delta-q^2\Delta$ where $q^2=\frac{M-\delta_m}{\Delta_m}$. It is straightforward to check that for  $\lvert \delta \rvert \leq \delta_m$ and $\lvert \Delta \rvert \leq \Delta_m$, $\dot{V}<0$ and $\lvert u \rvert \leq M$, whence concluding the locking section. To summarize, in this section we proved the existence of $\delta_{loc}$, such that if the sensor noise is bounded by $\delta_{loc}$ then the agent is able to lock onto an obstacle.\\
Note that tracking the nominal curve in the linear intermediate stage of locking is analogous to tracking the attraction line in absence of obstacles. Thus, we will use the same control law in order for the agent follow to the attraction line.

\subsection{Unlocking}\label{sec:unlocking}
As described in the algorithm, we will stop the tracking phase and begin the unlocking phase when the agent's position is in a neighbourhood the attraction line. We aim to drive the system to have velocity parallel to $v(0)$. As in locking, we define a local coordinate axis where the vertical axis is parallel to the velocity of the agent at the time instance when tracking stops and the horizontal axis is orthogonal to the later and points away from the obstacle. In the absence of noise, we define nominal trajectories,
\begin{subequations}
\begin{align}
\begin{split}
\begin{cases}
\overline{x}(t)=\frac{r_0}{\overline{u}}\left(1-{\rm e}^{-j\frac{a_0\overline{u}}{v_0}t}\right)\\
\overline{v}(t)=jv_0{\rm e}^{-j\left(\frac{a_0\overline{u}}{v_0}t\right)}\\
\overline{u}(t)=\frac{M}{\left(1+\epsilon\right)}
\end{cases}
\hspace{2.8cm}
t\in [0,t_1]
\end{split}\\
\begin{split}
\begin{cases}
\overline{x}(t)=\left(c_1+jc_2\right)+\frac{r_0}{\overline{u}}{\rm e}^{j\frac{a_0\overline{u}}{v_0}(t-T)+\theta}\\
\overline{v}(t)=jv_0{\rm e}^{j\frac{a_0\overline{u}}{v_0}(t-T)+\theta}\\
\overline{u}(t)=\frac{M}{\left(1+\epsilon\right)}
\end{cases}
\hspace{1cm}t\in (t_1,T]
\end{split}
\end{align}
\end{subequations}
where $t_1,t_2,T$ ensure that the above two circles are tangent to each other, $\cos{\theta}=\frac{\text{Re}\left(\hat{v}(0)\right)}{v_0}$ where $\hat{v}(0)$ is the initial agent velocity in the local unlocking coordinate frame, $\epsilon=0.151$ obtained from locking, $r=\frac{r_0(1+\epsilon)}{M}$, $c_1=\lambda\cos{\theta}-r\sin{\theta}$, $c_2=\lambda\sin{\theta}+r\cos{\theta}$ and
$\lambda=r\left(\cos{\theta}+\sqrt{\left(1-\sin{\theta}\right)\left(3+\sin{\theta}\right)}\right)$. When considering noise, we note that the unlocking control problem is analogous to the locking one (in unlocking there is no linear segment in the nominal trajectory) so the latter discussion applies. Just as in locking we have thus proved the existence of a sensor noise bound $\delta_{unloc}$ which guarantees unlocking.\\
Furthermore, note that the unlocking manoeuvre requires $r_0\left(\sqrt{3}+1\right)\frac{1+\epsilon}{M}$ of free-space (consider $\theta=0$ in the nominal trajectories to obtain the latter). However, we stress that such a manoeuvre is done while directing the sensor towards the obstacle (i.e. not looking ahead) so we do not expect the sensor to be aligned with the velocity once $v$ is the neighbourhood of $v(0)$. The worst case scenario occurs when we need to rotate the sensor by $\pi$ which requires a guaranteed free-space of $\frac{\pi v_0}{\gamma}$. Combining, the free-space and sensor considerations for the unlocking phase, we find that the free-neighbourhood required for unlocking is contained in the one required for locking. 
\end{document}